\documentclass[10pt,conference]{IEEEtran}
\IEEEoverridecommandlockouts

\usepackage{cite}
\usepackage{amsmath,amssymb,amsfonts}
\usepackage{algorithm}
\usepackage{algpseudocode}
\usepackage{textcomp}
\usepackage{xcolor}
\usepackage{enumitem}
\usepackage{listings}
\usepackage{float}
\usepackage{graphicx}
\usepackage{enumitem}
\usepackage{adjustbox}
\usepackage{tcolorbox}
\usepackage{svg}
\usepackage{balance}
\usepackage{booktabs}
\usepackage{url}
\usepackage{algorithm}
\usepackage{algpseudocode}
\usepackage{graphicx,calc}
\usepackage{textcomp}
\usepackage{xcolor}
\usepackage{listings}
\usepackage[utf8]{inputenc}
\usepackage[flushleft]{threeparttable}
\usepackage{balance}
\usepackage{wrapfig}
\usepackage{textcomp}
\usepackage{mathtools,amsthm}
\usepackage{tcolorbox}
\usepackage{xstring}
\usepackage{comment}
\usepackage{multicol}
\usepackage{tikz}
\usetikzlibrary{calc,arrows.meta}
\usepackage{xspace}
\usepackage{tcolorbox}
\usepackage{booktabs}

\newlength\myheight
\newlength\mydepth
\settototalheight\myheight{Xygp}
\newcommand{\bettersim}{{\raise.17ex\hbox{$\scriptstyle\sim$}}}

\usepackage{enumitem}
\newcommand{\squishlist}{
	\begin{itemize}[noitemsep, nolistsep, leftmargin=*]
		\setlength{\itemsep}{-0pt}
	}
	\newcommand{\squishstart}{
		\begin{itemize}
		}
		\newcommand{\squishend}{
		\end{itemize}
	}
	\newcommand{\PP}[1]{
		\vspace{3px}
		\noindent{\bf \textsc{\IfEndWith{#1}{.}{#1}{#1.}}}
	}
	\newcommand{\PPP}[1]{
		\vspace{2px}
		\indent{\it \IfEndWith{#1}{.}{#1}{#1.}}
	}
	
	\newcommand{\commentt}[1]{}

	\graphicspath{{images/}}

\newcommand{\smalltt}[1]{{\footnotesize\texttt{#1}}}
\newenvironment{rqanswer}{%
    \begin{tcolorbox}[colframe=black!90,colback=gray!10,boxrule=.75pt]
}{%
    \end{tcolorbox}
}
\newlength{\textfloatsepsave}
\usepackage{listings}
\usepackage{xcolor}

\colorlet{punctcolor}{red!60!black}
\colorlet{desccolor}{green!60!black}
\colorlet{altcolor}{blue!60!black}
\colorlet{kwcolor}{teal!60!black}
\definecolor{keywordcolor}{rgb}{0.13, 0.29, 0.53}

\begin{document}

\title{A Multi-Agent Approach for REST API Testing with Semantic Graphs and LLM-Driven Inputs}

\author{\IEEEauthorblockN{
Myeongsoo Kim\IEEEauthorrefmark{1}\textsuperscript{$1$}\thanks{\textsuperscript{$1$}Also with AWS AI Labs, Santa Clara, CA, USA; this work was performed before the author joined AWS AI Labs.},
Tyler Stennett\IEEEauthorrefmark{1}\textsuperscript{$2$},
Saurabh Sinha\IEEEauthorrefmark{2}\textsuperscript{$3$}, 
Alessandro Orso\IEEEauthorrefmark{1}\textsuperscript{$4$}}
\IEEEauthorblockA{
\IEEEauthorrefmark{1}Georgia Institute of Technology, Atlanta, GA, 30332, USA.\\
\IEEEauthorrefmark{2}IBM T.J. Watson Research Center, Yorktown Heights, NY, 10598, USA.\\
\{{\textsuperscript{$1$}mkim754, \textsuperscript{$2$}tstennett3\}@gatech.edu, \textsuperscript{$3$}sinhas@us.ibm.com, \textsuperscript{$4$}orso@cc.gatech.edu}}
}

\maketitle

\begin{abstract}
As modern web services increasingly rely on REST APIs, their thorough testing has become crucial. Furthermore, the advent of REST API documentation languages, such as the OpenAPI Specification, has led to the emergence of many black-box REST API testing tools. However, these tools often focus on individual test elements in isolation (e.g., APIs, parameters, values), resulting in lower coverage and less effectiveness in fault detection. To address these limitations, we present AutoRestTest, the first black-box tool to adopt a dependency-embedded multi-agent approach for REST API testing that integrates multi-agent reinforcement learning (MARL) with a semantic property dependency graph (SPDG) and Large Language Models (LLMs). Our approach treats REST API testing as a separable problem, where four agents---API, dependency, parameter, and value agents---collaborate to optimize API exploration. LLMs handle domain-specific value generation, the SPDG model simplifies the search space for dependencies using a similarity score between API operations, and MARL dynamically optimizes the agents' behavior. Our evaluation of AutoRestTest on 12 real-world REST services shows that it outperforms the four leading black-box REST API testing tools, including those assisted by RESTGPT (which generates realistic test inputs using LLMs), in terms of code coverage, operation coverage, and fault detection. Notably, AutoRestTest is the only tool able to trigger an internal server error in the Spotify service. Our ablation study illustrates that each component of AutoRestTest---the SPDG, the LLM, and the agent-learning mechanism---contributes to its overall effectiveness.
\end{abstract}


\section{Introduction}

Modern web services increasingly depend on REST (Representational State Transfer) APIs for efficient communication and data exchange~\cite{fielding2000architectural}. These APIs enable seamless interactions between various software systems using standard web protocols~\cite{pautasso2008}. REST APIs function through common internet methods and a design that allows the server and client to operate independently~\cite{richardson2013restful}. The prevalence of REST APIs is evident from platforms such as APIs Guru~\cite{apis_guru} and Rapid API~\cite{rapidapi}, which host extensive collections of API specifications.

In recent years, various automated testing tools for REST APIs have been developed (e.g.,~\cite{Corradini2022, atlidakis2019restler, karlsson2020quickrest, martin2021restest, karlsson2020automatic, zac2022schemathesis, wu2022combinatorial, kim2023reinforcement, tcases, liu2022morest,corradini2023automated,corradini2024rest,arcuri2025tool,le2024kat,chen2024dyner,stennett2025autoresttesttoolautomatedrest,kim2025llamaresttesteffectiverestapi}). These tools follow a sequential process: select an operation to test, identify operations that depend on the selected operation, determine parameter combinations, and assign values to those parameters. Feedback from response status codes is then used to adjust the exploration policy at each step, either encouraging or penalizing specific choices. Although significant research has been dedicated to optimizing each individual step---operation selection, dependency identification, parameter selection, and value generation---these tools treat each step in isolation, rather than as part of a coordinated testing strategy. This isolated approach can result in suboptimal testing with a high number of invalid requests. For instance, one endpoint might benefit from extensive exploration of different parameter combinations, whereas another endpoint might require focused exploration of previously successful parameter combinations with diverse input values.

Consequently, existing tools often achieve low coverage, especially on large REST services, as shown in recent studies (e.g., Language Tool, Genome Nexus, and Market in the ARAT-RL evaluation~\cite{kim2023reinforcement}, and Spotify and OhSome in the NLP2REST evaluation~\cite{kim2023enhancing}).

To overcome these limitations, we propose AutoRestTest, a new approach that integrates a semantic property dependency graph (SPDG) and multi-agent reinforcement learning (MARL) with Large Language Models (LLMs) to enhance REST API testing. Instead of traversing all operations to find dependencies by matching input/output properties, AutoRestTest uses an embedded graph that prioritizes the properties by calculating the cosine similarity between input and output names. 

Specifically, AutoRestTest employs four specialized agents to optimize the testing process. The \textit{dependency agent} manages and utilizes the dependencies between operations identified in the SPDG, guiding the selection of dependent operations for API requests. The \textit{operation agent} selects the next API operation to test, prioritizing operations likely to yield valuable test results based on previous results, such as successfully processed dependent operations. The \textit{parameter agent} chooses parameter combinations for the selected operation to explore different configurations. Finally, the \textit{value agent} manages the generation of parameter values using three data sources: values from dependent operations, LLM-generated values that satisfy specific constraints using few-shot prompting~\cite{brown2020language}, and type-based random values. The value agent learns which data source is most effective for each parameter type and context.

The AutoRestTest agents collaborate to optimize the testing process using the multi-agent value decomposition Q-learning approach~\cite{zhang2021multi, sunehag2017value}. When selecting actions, each agent is employed independently using the epsilon-greedy strategy for exploitation-exploration balancing. However, during policy updates using the Q-learning equation, AutoRestTest uses value decomposition to consider the joint actions across all agents. Through centralized policy updates, each agent converges toward selecting the optimal action while accounting for the actions of the other agents.

We evaluated AutoRestTest on 12 real-world RESTful services used in previous studies~\cite{kim2023enhancing,kim2023reinforcement}, including well-known services such as Spotify, and compared its performance with that of four state-of-the-art REST testing tools, recognized as top-performing tools in recent studies~\cite{kim2022automated,kim2023reinforcement,liu2022morest,zhang2022open}: RESTler~\cite{atlidakis2019restler}, EvoMaster~\cite{arcuri2019restful}, MoRest~\cite{liu2022morest}, and ARAT-RL~\cite{kim2023reinforcement}. To ensure a fair comparison, we used enhanced API specifications generated by RESTGPT~\cite{kim2023leveraging}, which augments REST API documents with realistic input values generated using LLMs. To measure effectiveness, we used code coverage for the open-source services, operation coverage for the online services, and fault detection ability---measured as internal server errors triggered---for all the services. These are the most commonly used metrics in this field~\cite{golmohammadi2023survey,kim2022automated}.

AutoRestTest demonstrated superior performance across all coverage metrics compared to existing tools. It achieved 58.3\% method coverage, 32.1\% branch coverage, and 58.3\% line coverage, significantly outperforming ARAT-RL, EvoMaster, RESTler, and MoRest by margins of 12--27\%. For closed-source services, AutoRestTest processed 25 API operations, compared to 9--11 operations processed by the other tools. These results show that AutoRestTest can perform more comprehensive API testing than the other tools. 

In terms of fault detection, AutoRestTest identified 42 operations with internal server errors, outperforming ARAT-RL (33), EvoMaster (20), MoRest (20), and RESTler (14). Notably, AutoRestTest was the only tool that detected an internal server error for Spotify~\cite{spotifyerror}. We reported the errors detected on FDIC, OhSome, and Spotify, as these are actively maintained projects. The OhSome error has been confirmed and fixed~\cite{ohsomeerror}; we are still waiting for feedback from the developers for the other bug reports.

We also performed an ablation study, which revealed the importance of AutoRestTest's key components. Removing temporal-difference Q-learning caused the largest performance drop, with method, line, and branch coverage falling to 45.6\% (-12.7), 18.2\% (-13.9), and 45.8\% (-12.5), respectively. SPDG removal reduced coverage to 46.7\% (-11.6), 18.7\% (-13.4), and 47.6\% (-10.7), while LLM removal led to 47.4\% (-10.9), 19.3\% (-12.8), and 45.8\% (-12.5). Overall, removing any component decreased coverage by 10.7--13.9\%, demonstrating the significance of each component in contributing to AutoRestTest's effectiveness.

The main contributions of this work are:
\begin{itemize}
\item A novel REST API testing technique that reduces the operation dependency search space using a similarity-based graph model, employs multi-agent reinforcement learning to consider optimization among the testing steps, and leverages LLMs to generate realistic test inputs.
\item Empirical results showing that AutoRestTest outperforms state-of-the-art REST API testing tools---even when provided with enhanced API specifications---by covering more operations, achieving higher code coverage, and triggering more failures.
\item An artifact including the AutoRestTest tool, the benchmark services used in the evaluation, and detailed empirical results, which serves as a comprehensive resource for supporting further research and replication of our results~\cite{artifact}.
\end{itemize}

\section{Background and Motivating Example}
\label{sec:background}

\begin{figure}[t]
  \centering
    \begin{lstlisting}
/register:
  post:
    tags:
      - customer-rest-controller
    summary: createCustomer
    operationId: createCustomerUsingPOST
    requestBody:
      description: user
      content:
        application/json:
          schema:
            type: object
            properties:
              links:
                type: array
                items:
                  $ref: '#/components/schemas/Link'
              email:
                type: string
                maxLength: 50
                pattern: "^[\\w-]+(\\.[\\w-]+)*@([\\w-]+\\.)+[a-zA-Z]+$"
              name:
                type: string
                maxLength: 50
                pattern: "^[\\pL '-]+$"
              password:
                type: string
                maxLength: 50
                minLength: 6
                pattern: "^[a-zA-Z0-9]+$"
    responses:
      "201":
        description: Created
        content:
          '*/*':
            schema:
              $ref: '#/components/schemas/UserDTORes'
      "401":
        description: Unauthorized
      "403":
        description: Forbidden
      "404":
        description: Not Found
    \end{lstlisting}
  \caption{A part of Market API's OpenAPI Specification.}
  \label{fig:market_oas}
\end{figure}

\subsection{REST APIs}
REST APIs are a type of web APIs that conform to RESTful architectural principles~\cite{fielding2000architectural}. These APIs enable data exchange between client and server through established protocols like HTTP~\cite{berners1996hypertext}. The foundation of REST lies in several key concepts: statelessness, cacheability, and a uniform interface~\cite{richardson2013restful}.

In RESTful services, clients interact with web resources by making HTTP requests. These resources represent various types of data that a client might wish to create, read, update, or delete. The API endpoint, defined by a specific resource path (e.g., /users), represents the resource that clients interact with. Different HTTP methods (e.g., \textsc{post}, \textsc{get}, \textsc{put}, \textsc{delete}) determine what actions can be performed on that resource, such as creating, reading, updating, or deleting data. Each unique combination of an endpoint's resource path and an HTTP method constitutes an operation (e.g., \textsc{get} /users or \textsc{post} /users). 

When a web service processes a request, it responds with headers, a body (if applicable), and an HTTP status code indicating the result. Successful operations typically return 2xx codes, while 4xx codes denote client-side errors, and a 500 status code indicates an internal server error. 

\subsection{OpenAPI Specification}

The OpenAPI Specification (OAS)~\cite{openapi}, which evolved from Swagger~\cite{swagger} in version 2 to the OAS in version 3, is a crucial standard for RESTful API design and documentation. As an industry-standard format, OAS defines the structure, functionality, and anticipated behaviors of APIs in a standardized and human-accessible way.

Figure~\ref{fig:market_oas} presents a portion of the Market API's OAS. This example highlights the structured approach of OAS in defining API operations. For instance, it specifies a \textsc{post} request on the \smalltt{/register} endpoint for creating a new customer. The request body is expected to be in \smalltt{application/json} format and must include properties such as \smalltt{links}, \smalltt{email}, \smalltt{name}, and \smalltt{password}, each with specific validation constraints. The responses section defines possible outcomes, including a \smalltt{201 Created} status for a successful request, along with \smalltt{401 Unauthorized}, \smalltt{403 Forbidden}, and \smalltt{404 Not Found} status codes for various error conditions.

\subsection{Large Language Models}

Large Language Models (LLMs), like the Generative Pre-trained Transformer (GPT) series, are at the forefront of advancements in natural language processing (NLP)~\cite{hadi2023large}. LLMs, trained on extensive text collections, can understand, interpret, and generate human-like text~\cite{openai2023gpt4}. The GPT series, including GPT-3, exemplifies these advanced LLMs, demonstrating a remarkable ability to produce human-like text for various applications, from education to customer service~\cite{brown2020language, baidoo2023education, cotton2023chatting}. Their versatility in handling diverse language-related tasks, from writing coherent text and translating languages to test case generation~\cite{pan2024multi}, highlights their advanced capabilities in human-like understanding~\cite{openai2023gpt4}.

\subsection{Multi-Agent Reinforcement Learning}

Reinforcement learning (RL) is a branch of machine learning where an agent learns to make decisions by interacting with its environment~\cite{sutton2018reinforcement}. In this process, the agent selects actions in various situations (states), observes the outcomes (rewards), and learns to choose the best actions to maximize the cumulative reward over time. RL involves a trial-and-error approach, where the agent discovers optimal actions by experimenting with different options and adjusting its strategy based on the observed rewards. Additionally, the agent must balance between exploring new actions to gain more knowledge and exploiting known actions that provide the best reward based on its current understanding. This balance between exploration and exploitation is often controlled by parameters, such as $\epsilon$ in the $\epsilon$-greedy strategy~\cite{sutton2018reinforcement}.

Multi-agent reinforcement learning (MARL) is an advanced extension of reinforcement learning that involves multiple agents interacting in a shared environment to achieve individual or collective goals~\cite{zhang2021multi}. MARL addresses the complexities of agent interactions, including cooperation, competition, and communication~\cite{hernandez2019survey}. Techniques such as cooperative learning, competitive learning, and communication and coordination methods enable agents to develop optimal strategies. Applications of MARL span various domains, including autonomous driving and robotic coordination, where agents must work together or compete to optimize their performance~\cite{shalev2016safe, busoniu2008comprehensive, silver2016mastering}. MARL is expected to be used in areas requiring complex multi-agent decision making~\cite{yang2020overview}.

\subsection{Motivating Example}

Figure~\ref{fig:market_oas} depicts the \smalltt{/register} endpoint in the Market API's OpenAPI specification. This endpoint is vital for creating user credentials needed in other components of the API. Existing REST API testing tools struggle to generate valid requests for this operation due to the strict parameter requirements: \smalltt{email}, \smalltt{name}, and \smalltt{password} are required parameters, each with specific constraints on valid values, while \smalltt{links} is an optional parameter. Moreover, these tools fail to prioritize information gained from a successful operation invocation in subsequent requests.

AutoRestTest addresses these issues through a multi-agent approach. The parameter agent employs reinforcement learning to identify valid parameter combinations, learning that (\smalltt{email}, \smalltt{name}, \smalltt{password}) is a required parameter set while avoiding invalid combinations that would trigger 400 errors. For these parameters, the value agent first determines the appropriate data source for each parameter---choosing between LLMs, dependency values from previous operations, or random values---because different parameters require different generation strategies to create valid values. For the \smalltt{/register} endpoint, it selects LLM generation as the parameters require specific formats (e.g., email format, password rules) that basic defaults cannot match, and dependency values are not available for this initial operation. It then generates context-aware values that comply with the specification's validation patterns. After a successful registration, the dependency agent utilizes the SPDG to identify operations that depend on user credentials (e.g., \smalltt{/customer/cart}, \smalltt{/customer/orders}) and propagates the registered user's information to these dependent operations. The operation agent then prioritizes testing these dependent operations, while the parameter and value agents reuse the strategies that were successful for the \smalltt{/register} endpoint. Through value decomposition, MARL enables efficient policy updates by appropriately distributing rewards to each agent based on their contribution.

\section{Our Approach}
\label{sec:our_approach}


\begin{figure*}[ht]
\centering
\includegraphics[width=.9\textwidth]{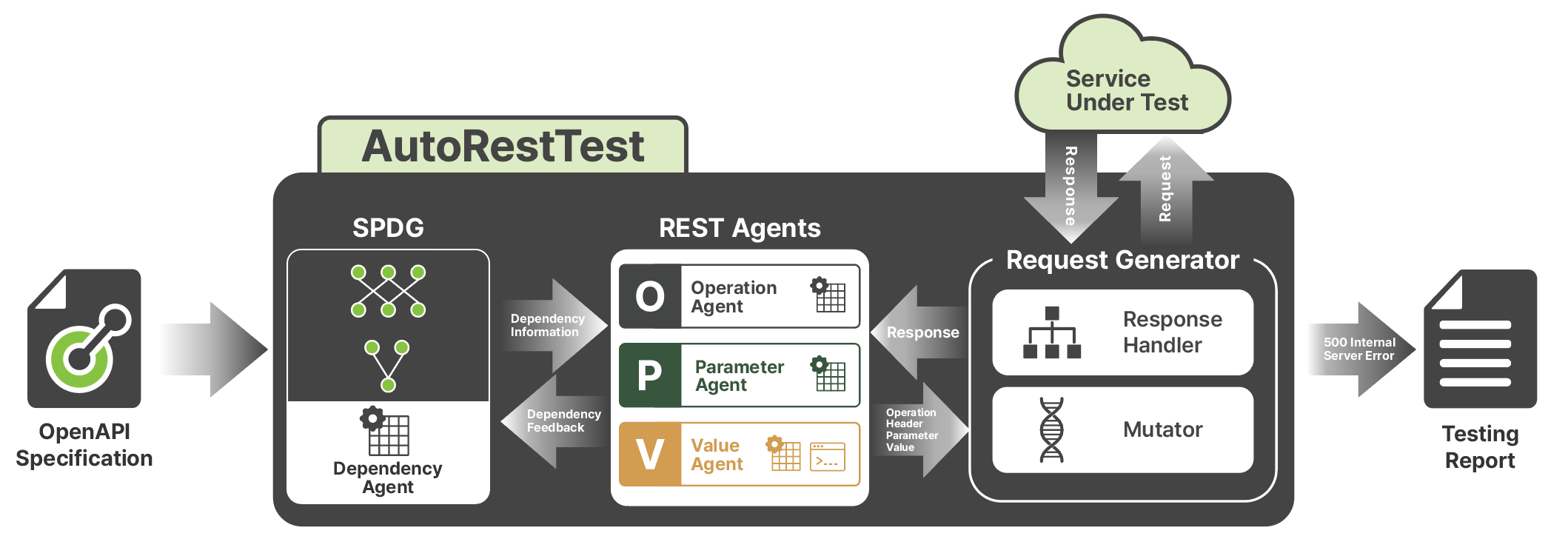}
\vspace*{-8pt}
\caption{Overview of our approach.}
\vspace*{-15pt}
\label{fig:overview}
\end{figure*}

\subsection{Overall Workflow}
\label{sec:approach-overview}

Figure~\ref{fig:overview} illustrates the architecture of AutoRestTest, highlighting its core components: the SPDG, the REST agents, and the request generator. The overall workflow consists mainly of two phases: initialization and testing execution.

\subsubsection{Initialization Phase}
The process begins with parsing the OpenAPI specification to extract endpoint information, parameters, and request/response schemas. Using the parsed specification, the dependency agent constructs the SPDG as a directed graph in which nodes represent API operations and edges represent potential dependencies between operations. Each weighted edge $e = (a, b)$ indicates that operation $b$ provides values that can be used by operation $a$ (i.e., $a$ depends on $b$), with the edge weight (a value between 0 and 1) representing the semantic similarity between the operations' inputs and outputs (see Section~\ref{sec:spdg}).
These initial dependencies are later validated and refined through the testing process based on actual server responses. 

The REST agents (operation, parameter, value, and dependency agents) are then initialized with their respective Q-tables. Each agent serves a specific purpose in the testing process: the operation agent selects API operations to test, the parameter agent determines parameter combinations, the value agent generates appropriate parameter values for each parameter, and the dependency agent manages operation dependencies from the SPDG.

\subsubsection{Testing Execution Phase}

The testing execution follows an iterative process. In each iteration, the operation agent first selects the next API operation based on its learned Q-values and exploration strategy. Next, the parameter agent determines which parameters to include, considering both required and optional parameters from the specification. The value agent then generates parameter values using dependencies identified by the dependency agent, LLM-generated values, or default assignments for basic parameter types. Using the SPDG, the dependency agent identifies any dependencies between the selected parameters and those used in previous operations. Finally, the request generator constructs the API request, with the mutator component modifying 20\% of requests to test error handling and trigger potential 500 response codes, similar to prior work~\cite{kim2023reinforcement}.

Once the request is sent to the Service Under Test (SUT) and a response is received, the response is used to update the Q-tables of all agents through reinforcement learning, refine SPDG dependencies, and store any 500 responses for the final testing report. This cycle continues until the testing time budget is exhausted. The SPDG refinement process involves increasing or decreasing edge weights, driven by rewards and penalties for dependencies, based on server responses---successful dependencies (validated by 2xx response codes) increase edge weights, whereas failed dependencies reduce edge weights, with heavily penalized edges being effectively removed from consideration.
This continuous refinement helps ensure the accuracy of the dependency graph over time.

\subsection{Q-Learning and Agent Policy}
\label{sec:qlearning}

Both the SPDG and REST agent modules use the Q-learning algorithm~\cite{watkins1992q} with value decomposition within their respective agents. During initialization, each agent creates a Q-table data structure that maps available actions to their expected cumulative rewards. When request generation begins, AutoRestTest addresses the two primary components of Q-learning---action selection and policy optimization---to facilitate effective communication between agents.

\subsubsection{Action Selection}
\label{subsec:action_sel}

During action selection, agents independently choose between exploiting their best-known option or exploring new options randomly. To balance these choices, all agents follow an epsilon-greedy strategy: with probability \(\epsilon\), the agent selects a random action (exploration), and with probability \(1-\epsilon\), it selects the action with the highest Q-value (exploitation), likely yielding the best results.

AutoRestTest utilizes epsilon-decay to guarantee all actions are adequately explored in its initial stages. Starting with an epsilon value of 1.0, this value decreases linearly to 0.1 over the duration of the tool's operation. This strategy, commonly used in practice, has been shown to improve performance by balancing exploration and exploitation \cite{kumar2021importance}.

\subsubsection{Policy Optimization}

\begin{figure*}[t]
\centering
\includegraphics[width=.9\textwidth]{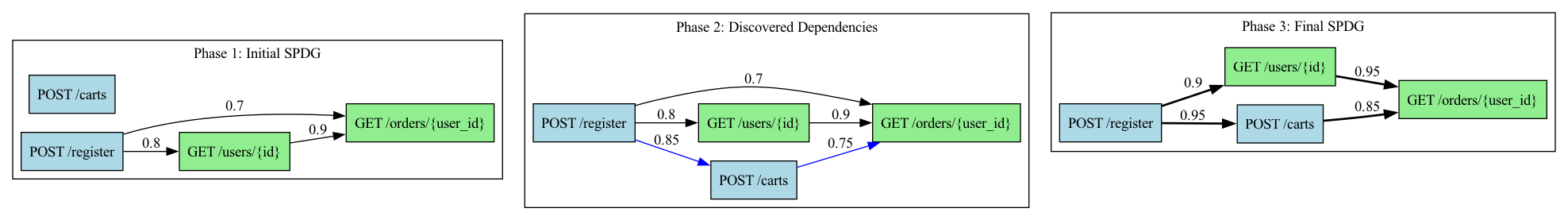}
\vspace*{-12pt}
\caption{Illustration of SPDG construction and refinement.}
\vspace*{-12pt}
\label{fig:spdg}
\end{figure*}

After receiving a response from the server, agents update their Q-table values using the temporal-difference update rule of the Q-learning algorithm, derived from the Bellman equation~\cite{sutton2018reinforcement}. This update aims to maximize the expected cumulative reward for each action taken. The Q-learning update rule is given by:
\begin{equation}
    Q(s,a) \leftarrow Q(s,a) + \alpha \delta
    \label{eq:q_update}
\end{equation}
where \( \delta \) represents the temporal-difference error, calculated as:
\begin{equation}
    \delta = r + \gamma \max_{a'} Q(s', a') - Q(s, a)
    \label{eq:td_error}
\end{equation}
where \( \alpha \) is the learning rate, \( \gamma \) is the discount factor, \( r \) is the received reward, \( s \) is the current state, \( s' \) is the next state, \( a \) is the current action, and \( a' \) is an action in state \( s' \). 

Given the complexity of the multi-agent environment, AutoRestTest leverages value decomposition to optimize the joint cumulative reward, which has shown improvements for policy acquisition over independent learning. This approach assumes that the joint Q-value can be decomposed additively as follows \cite{sunehag2017value}:
\begin{equation}
    Q(s,a) = \sum_i^n Q_i(s,a_i)
    \label{eq:value_decomp}
\end{equation}
where $Q_i$ and $a_i$ represent the Q-value and action for agent $i$ in shared state $s$.

The temporal-difference error, which measures the difference between current Q-values and the optimal target Q-values, can be redefined using value decomposition to reflect the displacement from the optimal joint Q-value:
\begin{equation}
    \delta = r + \gamma \max_{a'} \sum_i^n Q_i(s',a'_i) - \sum_i^n Q_i(s,a_i)
    \label{eq:joint_td_error}
\end{equation}

Using this redefined temporal-difference error, each agent updates its Q-table to converge towards the optimal joint Q-value, as depicted in the following temporal-difference equation:

\begin{equation}
    \begin{split}
    Q_i(s,a_i) \leftarrow Q_i(s,a_i) + \alpha \Big[ r + \gamma \max_{a'} \sum_{i=1}^n Q_i(s',a'_i) \\
    - \sum_{i=1}^n Q_i(s,a_i) \Big]
    \end{split}
    \label{eq:joint_q_update}
\end{equation}

This value decomposition approach enables each agent to select actions independently while maintaining centralized policy updates, simplifying the implementation while enhancing coordination across agents~\cite{sunehag2017value}.

For reward delegation, the dependency, value, and parameter agents are optimized to reward actions that generate 2xx status codes, whereas the operation agent rewards the selection of operations that generate 4xx and 5xx status codes. This balance in behavior is intended to maximize coverage by encouraging repeated attempts at creating successful requests for operations that frequently yield 4xx and 5xx status codes. For the hyperparameters $\alpha$ and $\gamma$, like ARAT-RL~\cite{kim2023reinforcement} and other related work~\cite{gibberblot,kim2024debloating}, we use values 0.1 and 0.9, respectively.

\subsection{Semantic Property Dependency Graph}
\label{sec:spdg}

The construction of the SPDG begins with parsing the OpenAPI specification to extract information about API endpoints, parameters, and request/response schemas. As shown in Figure~\ref{fig:spdg}, the SPDG is initialized as a directed graph where nodes represent API operations and edges represent potential dependencies between operations based on semantic similarities.

The initialization process involves two main steps. First, for each operation in the API specification, we create a node containing the operation's ID, parameters, and response schemas. Then, we identify potential dependencies between operations by computing semantic similarities between their parameter names (inputs) and response field names (outputs) using cosine similarity with pre-trained word embeddings (e.g., GloVe~\cite{pennington2014glove}). When comparing two operations, if their similarity score exceeds 0.7,\footnote{We selected this value based on previous studies~\cite{rekabsaz2017exploration,kim2023enhancing}.} we create an edge between them with the similarity score as its weight. To ensure all operations have some potential dependencies to explore, if an operation has no edges with scores above the threshold, we connect it to the five most similar operations.\footnote{Top-five is a common threshold in top-k similarity matching~\cite{komiya2013question,yao2023beyond,li2023using}.}

Phase 1 of Figure~\ref{fig:spdg} provides an example of the initial SPDG generated by our technique. For example, the edge between \textsc{get} \smalltt{/users/{id}} and \textsc{get} \smalltt{/orders/{user\_id}} has a weight of 0.9 due to the high similarity between the names of the inputs and outputs of these two operations.

\subsubsection{Dependency Agent}
\label{sec:dependency-agent}

The dependency agent manages and uses the dependencies between operations captured in the SPDG. It uses a Q-table to represent the validity of these dependencies, operating similarly to a weighted graph, where edges are assigned values that reflect confidence in each dependency. Specifically, the Q-table encodes edges from the SPDG, categorizing dependencies by parameter type (query or body) and target (parameters, body, or response). Higher Q-values on an edge indicate greater confidence in the reliability of that operation dependency, causing the agent to prioritize these relationships on future requests.

For each parameter and body property in a selected operation, the dependency agent refers to its Q-table to identify a dependent parameter, body, or response, as well as the associated operation ID. For example, as illustrated in Phase 3 of Figure~\ref{fig:spdg}, when testing \textsc{get} \smalltt{/orders/{user\_id}}, the agent might use the value for \smalltt{user\_id} from a successful \textsc{get} \smalltt{/users/{id}} response, reflected by the strengthened edge weight of 0.95. The dependency agent consults AutoRestTest's tables storing successful parameters, request body properties, and decomposed responses to ensure that the selected dependency has available values. AutoRestTest recursively deconstructs response objects, allowing the dependency agent to access nested properties within response collections. Required parameters without available dependencies are fulfilled using value mappings provided by the value agent.

Although the SPDG significantly reduces the search space for operation dependencies, the reliance on semantic similarity may overlook potential candidates. To address this, the dependency agent permits random dependency queries during exploration. As shown in Phase 2 of Figure~\ref{fig:spdg}, if a random dependency successfully generates a valid input, AutoRestTest identifies the contributing factor and adds this new edge to the SPDG. For instance, during testing, the agent might discover that \textsc{post} \smalltt{/carts} returns a \smalltt{cart\_id} that can be used as input for \textsc{get} \smalltt{/orders/{user\_id}}, leading to a new edge with weight 0.75. Similarly, when the specification contains undocumented response values (e.g., if \textsc{post} \smalltt{/register} returns additional user-related fields not specified in the OpenAPI document), the dependency agent evaluates these new properties for potential dependencies with other operations, as demonstrated by the edge between \textsc{post} \smalltt{/register} and \textsc{post} \smalltt{/carts} with weight 0.85 in Phase 2. 

\subsection{REST Agents}
\label{sec:rest-agents}

\subsubsection{Operation Agent}
\label{sec:operation-agent}

This agent is tasked with selecting the next API operation to test. It employs reinforcement learning to prioritize operations that are likely to yield meaningful test results based on prior experiences. 
This agent uses a simplified state model that only tracks whether an operation is available for selection. The agent's action space encompasses all possible operations in the API specification. Each operation's Q-value is initialized to 0 and is updated based on response codes from the server. The Q-table for the operation agent stores cumulative rewards representing the proportion of unsuccessful requests for each operation in the provided service, as discussed in greater detail in the next paragraph.

The operation agent acts as the forerunner of the testing process, selecting an operation that dictates the state of the remaining agents. While the remaining agents coordinate to generate valid test cases for a given operation, the operation agent is tasked with identifying unsuccessful operations for retesting. Consequently, while the remaining agents update their Q-value using the value decomposition temporal-difference equation (Equation~\ref{eq:joint_q_update} in Section~\ref{sec:qlearning}), the operation agent updates its Q-values independently using a structured reward system: +2 for server errors (5xx), +1 for client errors (4xx, excluding 401 and 405), -1 for successful responses (2xx), -3 for authentication failures (401), and -10 for invalid methods (405). This reward structure encourages the agent to prioritize exploring problematic endpoints while severely penalizing systematically invalid requests and mildly penalizing endpoints with consistently successful requests.

As an example, consider the customer registration endpoint in the Market API (Figure~\ref{fig:market_oas}). The Q-value for the endpoint is initially 0. After initial attempts at processing the operation fail, the Q-table value might increase (e.g., to 1), prompting the agent to prioritize further testing on this endpoint. Conversely, a successful test case would decrease the Q-value, directing the agent to explore other (challenging) endpoints.

\subsubsection{Parameter Agent}
\label{sec:parameter-agent}

The parameter agent is responsible for selecting parameters for the chosen API operation. It ensures that parameters used across requests are both valid and varied, covering a range of testing scenarios while addressing inter-parameter dependencies. For each operation, the parameter agent initializes a state containing the operation ID, available parameters, and required parameters, and defines its action space as possible parameter combinations. The Q-values associated with each state-action pair are initialized to 0.

Consider again the Market API's customer registration endpoint shown in Figure~\ref{fig:market_oas}. The parameter agent initializes with the following state: $s =$ \{\smalltt{createCustomerUsingPOST}, [\smalltt{email}, \smalltt{name}, \smalltt{password}, \smalltt{links}], [\smalltt{email}, \smalltt{name}, \smalltt{password}]\}, where the first list contains all available parameters from the request body schema, whereas the second list contains only the required parameters according to the endpoint's specification. The agent initializes a Q-table for this operation, mapping various parameter combinations to Q-values. 

This setup ensures that different parameter combinations (limited to 10 by default, to account for space restrictions) are sufficiently represented in the agent's action space. The agent updates its Q-values using the value decomposition temporal-difference equation (Equation~\ref{eq:joint_q_update} in Section~\ref{sec:qlearning}) and the following reward structure: -1 for server errors (5xx), -2 for client errors (4xx), and +2 for successful responses (2xx).

Importantly, unused parameter combinations receive a neutral update (effectively 0) in the Q-learning process, maintaining their initial Q-values. These unused combinations are prioritized over combinations with negative rewards and deprioritized relative to those with positive rewards. For example, for the registration endpoint, if the parameter combination (\smalltt{email}, \smalltt{name}, \smalltt{password}) consistently yields positive rewards, the unused combination (\smalltt{email}, \smalltt{name}, \smalltt{password}, \smalltt{links}) retains its initial Q-value and may be selected during exploration to test scenarios with optional parameters. 
Suppose that the Q-values for these parameter combinations evolved to 0.8 and 0.3, respectively. This would suggest that the inclusion of the \smalltt{links} parameter is problematic and result in a lower Q-value for the configuration with all parameters.

Through this reward scheme, AutoRestTest effectively identifies valid parameter sets and addresses challenges related to undocumented inter-parameter dependency requirements, particularly in complex scenarios such as user registration, where certain parameters must be present and correctly formatted.

\subsubsection{Value Agent}
\label{sec:value-agent}

This agent is responsible for generating and assigning values to parameters selected by the parameter agent. For each parameter of an operation, it maintains a state containing the operation ID, parameter name, parameter type, and OpenAPI schema constraints, with its actions corresponding to possible data sources for parameter value assignment. The Q-values for each state-action pair are initialized to 0.

Consider the Market API's customer registration endpoint in Figure~\ref{fig:market_oas}. The value agent initializes the following states for \smalltt{email}, \smalltt{name}, and \smalltt{password} parameters:
\begin{itemize}[left=0pt]
\small
\item $s =$ \{\smalltt{createCustomerUsingPOST},~\smalltt{email}, \smalltt{string}, \{pattern: 
{\footnotesize \verb|^[\w-]+(\.[\w-]+)*@([\w-]+\.)+[a-zA-Z]+$|}
\}\}
\item $s =$ \{\smalltt{createCustomerUsingPOST}, \smalltt{name}, \smalltt{string}, \{pattern: {\footnotesize \verb|^[\pL '-]+\$|}, maxLength: 50\}\}
\item \{\smalltt{createCustomerUsingPOST}, \smalltt{password}, \smalltt{string}, \{pattern: {\footnotesize \verb|^[a-zA-Z0-9]+$|}, minLength: 6, maxLength: 50\}\}
\end{itemize}

To generate a diverse set of inputs, the value agent can select inputs from the following data sources:

\begin{itemize}[left=0pt]

\item \textit{Operation Dependency Values:} When selected, the value agent collaborates with the dependency agent to map dependent values to the selected parameter. For example, the registration endpoint might reuse email addresses from prior successful registrations to test duplicate user scenarios.

\item \textit{LLM Values}: For this data source, the value agent creates (or parses if already created) values using few-shot prompting with LLMs~\cite{brown2020language}.\footnote{In this work, we used GPT-3.5 Turbo with a temperature setting of 0.8 because of its known performance in REST API contexts~\cite{kim2023leveraging}.} For instance, the LLM might generate ``john.doe@example.com" as the input value for the \smalltt{email} parameter based on the specified pattern.

\item \textit{Random Values}: When this option is selected, the value agent generates random values based on the type of the selected parameter. For example, it may create a random sequence of 1-50 characters for strings, a random number between -1024 and 1024 for integers, and a random true/false value for boolean types.

\end{itemize}

Once a request is completed with values from the chosen data source, the agent updates its Q-values based on the temporal-difference equation (Equation~\ref{eq:joint_q_update} in Section~\ref{sec:qlearning}), using the same reward strategy as the parameter agent (\S\ref{sec:parameter-agent}) to refine its value generation.

For the registration endpoint, for instance, the Q-values across parameters show an average of 0.5 for LLM values, 0.2 for random values, and -0.7 for operation dependency values.
In analyzing these Q-values, we observe that because the registration endpoint is required for account creation, it is less likely to derive values from operation dependencies, which results in a lower Q-value for the dependency source. Although random values are effective for simpler parameters, such as \smalltt{name}, the LLM-generated values perform better for pattern-constrained fields, such as \smalltt{email} and \smalltt{password}.

\subsection{Request Generator}
\label{sec:request-generator}

The request generator constructs and dispatches API requests to the SUT. It works closely with the REST agents to ensure that the generated requests are both effective and comprehensive. Upon receiving responses from the SUT, the response handler processes these results and provides feedback to the REST agents, allowing refinement of future requests.

The mutator's purpose is to generate invalid requests to uncover unexpected behaviors (e.g., 500 responses). This is a crucial part of REST API testing frameworks, as state-of-the-art tools employ similar strategies such as mutating parameter types, values, and headers (e.g., using an invalid content type in the header). The mutator follows these conventions and mutates 20\% of the generated requests, a strategy used by the most recent tool~\cite{kim2023reinforcement}.

The request generator interacts with the REST agents to construct API requests, relying on them for detailed information about the operation to test, the parameters to use, and appropriate values for those parameters. Specifically:

\begin{itemize}[left=0pt]
\item The operation agent selects the API operation to test.
\item The parameter agent identifies and optimizes the parameters for the chosen operation.
\item The value agent generates realistic and effective values for the parameters.
\end{itemize}

Using this information, the request generator constructs a complete API request.
The request generator then dispatches the request to the SUT, which processes the request and returns a response. The response handler analyzes this response to detect any errors or unexpected behaviors. Insights from these analyses are fed back to both the REST agents and the dependency agent, allowing them to refine their strategies for future requests.

The interactions between the dependency module, the REST agents, the request generator, and the SUT establish a robust feedback loop that enhances the overall effectiveness of the testing process. This collaborative approach ensures that the generated requests are not only comprehensive but also tailored to uncover potential issues within REST APIs.

\section{Evaluation}
\label{sec:evaluation}

In this section, we present the results of empirical studies conducted to assess AutoRestTest. Our evaluation aims to address the following research questions:

\begin{enumerate}
\item \textbf{RQ1}: How does AutoRestTest compare with state-of-the-art REST API testing tools in terms of code coverage and operation coverage achieved?
\item \textbf{RQ2}: In terms of error detection, how does AutoRestTest perform in triggering 500 (Internal Server Error) responses compared to state-of-the-art REST API testing tools?
\item \textbf{RQ3}: How do the main components of AutoRestTest (MARL, SPDG, and LLM-based input generation) contribute to its overall performance?
\end{enumerate}

\subsection{Experiment Setup}
\label{sec:experiment_setup}

We conducted our experiments on two cloud VMs, each equipped with a 48-core Intel(R) Xeon(R) Platinum 8260 processor with 128 GB RAM. To ensure consistent test conditions, we restarted the services and restored their databases in each testing session to eliminate potential state dependency effects across sessions. We used the default configuration and database settings for each service. We allocated dedicated resources to each service and testing tool, running them sequentially to prevent interference. Throughout the experiments, we closely monitored CPU and memory usage to ensure optimal performance without encountering resource constraints.

\begin{table}[t]
    \centering
    \caption{REST services used in the evaluation.}
    \resizebox{.8\columnwidth}{!}{%
    \begin{tabular}{lrr}
        \toprule
        \textbf{REST Service} & \textbf{Lines of Code} & \textbf{\#Operations} \\ \midrule
        Features Service & 1688 & 18\\
        Language Tool & 113170 & 2 \\ 
        Rest Countries & 1619 & 22\\ 
        Genome Nexus & 22143 & 23\\ 
        Person Controller & 601 & 12\\ 
        User Management & 2805 & 22\\ 
        Market & 7945 & 13 \\ 
        Project Tracking System & 3784 & 67\\ 
        YouTube-Mock & 2422 & 1\\
        FDIC & -- & 6\\
        Spotify & -- & 12\\
        Ohsome API & -- & 122\\
        \bottomrule
    \end{tabular}
    }
    \label{tab:service_lines_of_code}
\end{table}

For our evaluation, we relied on the same set of REST API testing tools and services used by ARAT-RL~\cite{kim2023reinforcement}. Accordingly, we compared AutoRestTest with ARAT-RL~\cite{kim2023reinforcement}, EvoMaster~\cite{arcuri2019restful}, MoRest~\cite{liu2022morest}, and RESTler~\cite{atlidakis2019restler}. Specifically, we used the latest released version or the latest commit when a release was unavailable: RESTler v9.2.4, EvoMaster v3.0.0, ARAT-RL v0.1, MoRest (obtained directly from the authors). 

The ARAT-RL benchmark dataset has 10 RESTful services.
In addition to these services, we included the services from the RESTGPT study~\cite{kim2023leveraging}. Out of the total 16 services, we excluded SCS and NCS because they were written by EvoMaster's authors, and we aimed to avoid potential bias. We also excluded OCVN due to authentication issues. Lastly, we excluded OMDB, which is a toy online service with only one API operation that all testing tools can process in a second. Ultimately, we used 12 services: Features Service, Language Tool, REST Countries, Genome Nexus, Person Controller, User Management Microservice, Market Service, Project Tracking System, OhSome, YouTube-Mock, and Spotify. Table~\ref{tab:service_lines_of_code} lists the open-source services along with the lines of code and the number of API operations in each service.

For a fair comparison, because our tool utilizes LLM calls, we used the enhanced specification generated by RESTGPT, which adds realistic testing inputs to the specification using LLMs. Moreover, we used GPT-3.5-Turbo, as RESTGPT utilized this model. Based on a recent survey that describes settings and metrics for REST API testing~\cite{golmohammadi2023survey}, we used a one-hour time budget with ten repetitions to compute the results. To measure effectiveness and error-finding ability, we used code coverage (open-source services only), number of successfully processed operations in the specification, and number of 500 status codes, which are the most popular metrics in the literature. To collect code coverage, we used Jacoco~\cite{jacoco}. For the number of processed operations, we used the script from the NLP2REST repository~\cite{nlp2restartifact}. For the number of 500 status codes, we used the script available in the ARAT-RL repository~\cite{aratartifact}. This script collects 500 status codes by tracking the HTTP responses, and removes the duplicated 500 codes using the server response message for each operation.

\begin{figure*}[ht]
\centering
\includegraphics[width=15.3cm, height=10.2cm]{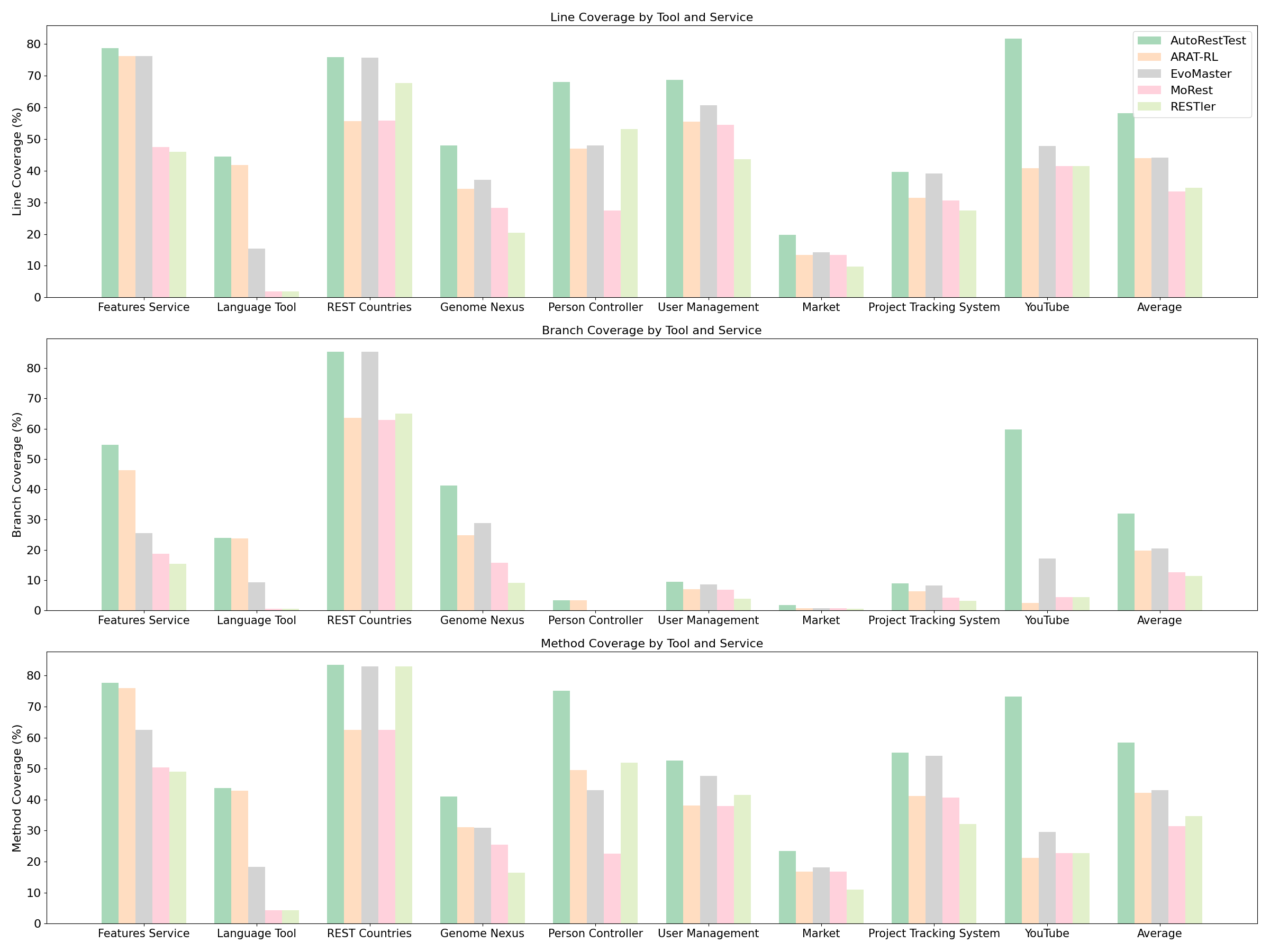}
\vspace*{-10pt}
\caption{Comparison of code coverage metrics across tools and services: line, branch, and method coverage.}
\vspace*{-10pt}
\label{fig:coverage}
\end{figure*}

\subsection{RQ1: Effectiveness}

The effectiveness of AutoRestTest is evaluated based on its ability to comprehensively cover more code compared to other tools. Figure~\ref{fig:coverage} illustrates the line, branch, and method coverage achieved by each testing tool on the nine open-source services in our benchmark; additionally, it shows the average coverage across these APIs.

As shown in Figure~\ref{fig:coverage}, AutoRestTest outperformed the other tools in terms of method coverage, achieving 58.3\% coverage on average, compared to ARAT-RL (42.1\%), EvoMaster (43.1\%), MoRest (31.5\%), and RESTler (34.7\%).
This represents a significant coverage gain ranging from 15.2 to 26.8 percentage points. Similarly, AutoRestTest achieved higher line coverage, 58.3\% on average, compared to the other tools, which achieved 44\%, 44.1\%, 33.4\%, and 34.6\%,  respectively. Finally, in terms of branch coverage, AutoRestTest again outperformed the other tools, achieving 32.1\% coverage on average compared to the other tools, which achieved 19.8\%, 20.5\%, 12.7\%, and 11.4\%, respectively.

In our evaluation, we also measured the number of processed operations for online services for which source code is unavailable: OhSome and Spotify. Table~\ref{tab:processed_operations} presents these results, which highlight AutoRestTest's ability to handle a larger number of operations. Specifically, AutoRestTest exercised 25 operations in total, compared to ARAT-RL, EvoMaster, MoRest, and RESTler, which processed 11, 10, 10, and 9 operations, respectively. These results on achieved code coverage and successfully exercised operations demonstrate AutoRestTest's effectiveness on a range of different REST APIs and how it improves on the state of the art.

\begin{table}[t]
\centering
\caption{Number of operations exercised.}
\resizebox{\columnwidth}{!}{%
\begin{tabular}{lccccc}
\toprule
& \textbf{AutoRestTest} & ARAT-RL & EvoMaster & MoRest & RESTler \\
\midrule
FDIC & 6 & 6 & 6 & 6 & 6 \\
OhSome & 12 & 0 & 0 & 0 & 0 \\
Spotify & 7 & 5 & 4 & 4 & 3 \\
\midrule
Total & \textbf{25} & 11 & 10 & 10 & 9 \\
\bottomrule
\end{tabular}
}
\label{tab:processed_operations}
\end{table}

In most cases, there were notable performance gains in Genome Nexus, Person Controller, User Management, Market, YouTube, OhSome, and Spotify. Conversely, for Features Service, REST Countries, and Project Tracking System, AutoRestTest's results did not show much difference compared to the second-best performing tool in our set. These four services have a notable characteristic in common: the number of input parameters in their APIs is mostly 1 to 2, and the services are therefore easier to test. This result shows that AutoRestTest can effectively explore REST APIs, especially for services with a large search space.

\begin{rqanswer}
AutoRestTest achieves considerable gains in code coverage, with method coverage increasing between 15.2 to 26.8 percentage points, line coverage between 14.2 and 24.8 percentage points, and branch coverage between 11.6 and 20.7 percentage points compared to the other tools considered. The improvement in performance is particularly noticeable on large and complex services with many input parameters.
\end{rqanswer}

\subsection{RQ2: Fault Detection Capability}

\begin{table}[t]
\centering
\caption{Service failures triggered (500 response codes).}
\resizebox{\columnwidth}{!}{%
\begin{tabular}{lccccc}
\toprule
REST APIs & \textbf{AutoRestTest} & ARAT-RL & EvoMaster & MoRest & RESTler \\
\midrule
Features Service & 1 & 1 & 1 & 1 & 1 \\
Language Tool & 1 & 1 & 1 & 0 & 0 \\
REST Countries & 1 & 1 & 1 & 1 & 1 \\
Genome Nexus & 1 & 1 & 0 & 1 & 0 \\
Person Contoller & 8 & 8 & 8 & 8 & 3 \\
User Management & 1 & 1 & 1 & 1 & 1 \\
Market & 1 & 1 & 1 & 1 & 1 \\
Project Tracking System & 1 & 1 & 1 & 1 & 1 \\
YouTube & 1 & 1 & 1 & 1 & 1 \\
FDIC & 6 & 6 & 6 & 6 & 6 \\
OhSome & 20 & 12 & 0 & 0 & 0 \\
Spotify & 1 & 0 & 0 & 0 & 0 \\
\midrule
Total & \textbf{42} & 33 & 20 & 20 & 14 \\
\bottomrule
\end{tabular}%
}

\label{tab:500errors}
\end{table}

We evaluated the fault detection capability of AutoRestTest by counting how many 500 Internal Server Errors it identified. Table~\ref{tab:500errors} shows the number of such errors detected by AutoRestTest, ARAT-RL, EvoMaster, MoRest, and RESTler. As the data in the table show, AutoRestTest detected a total of 42 500 Internal Server Errors across the evaluated REST APIs, far outperforming the other tools on this metric. ARAT-RL detected 33 errors, EvoMaster and MoRest detected 20 errors each, and RESTler detected 14 errors.

Specifically, AutoRestTest identified significantly more errors in the OhSome service (20 errors) compared to ARAT-RL (12 errors), with none detected by EvoMaster, MoRest, or RESTler. Additionally, AutoRestTest was the only tool to detect an error in the Spotify service. It is important to note that both OhSome and Spotify are active services; Spotify, for example, has 615 million users, and the OhSome service has recent GitHub commits. We reported the detected issues, and the OhSome errors were accepted, whereas we are still waiting for Spotify's response~\cite{spotifyerror,ohsomeerror}. This improvement in fault detection is somehow expected, as AutoRestTest achieves the highest coverage among the other tools, which is strongly correlated with fault-finding ability in REST API testing~\cite{kim2022automated}.

To illustrate the utility of AutoRestTest's specific components in fault detection on a specific example, consider the following sequence of operations in the Ohsome service, which shows the capabilities of the SPDG. AutoRestTest begins by successfully querying the \textsc{post} \smalltt{/elements/area/ratio} endpoint from the Ohsome service with its \smalltt{filter2} parameter assigned to \smalltt{node:relation}. Subsequently, AutoRestTest targets the \textsc{get} \smalltt{/users/count/groupBy/key} endpoint, where the dependency agent applies the SPDG's semantically-created dependency edges to identify a potential connection between the \smalltt{filter} parameter of the new operation and the \smalltt{filter2} parameter of the previous operation. When the dependency agent reuses the \smalltt{node:relation} value from the previously successful \smalltt{filter2} parameter in the new request, the SPDG uncovers an unexpected 500 Internal Server Error. Typically, an invalid \smalltt{filter} value would trigger a client error, but this server-side error indicates that the SPDG identified a deeper, unanticipated fault in the server's value handling. The other tools overlook the correlation between the two parameters due to either the minor naming differences or improper dependency modeling, and fail to expose this error.

For another example, AutoRestTest shows the effectiveness of its LLM value generation in the interactions with the Spotify API. The \textsc{get} \smalltt{/playlists/{playlist\_id}/tracks} operation in Spotify's API requires specific knowledge regarding Spotify's \smalltt{playlist\_id} formation. Spotify generates Spotify IDs for playlists that are typically 22 characters long with constraints on the permitted letters and patterns. Where many tools fail to create valid IDs, AutoRestTest's value agent leverages its LLM to generate valid Spotify IDs for playlists. AutoRestTest is thus able to successfully query the \textsc{get} \smalltt{/playlists/{playlist\_id}/tracks} operation, with rippling effects across the service. For instance, after retrieving the International Standard Recording Code (ISRC) from the playlist's tracks, AutoRestTest's mutator randomly selects an ISRC to use as the \smalltt{user\_id} in the subsequent \textsc{get} \smalltt{/users/{user\_id}/playlists} operation. This sequence reveals a hidden dependency conflict that results in a 500 Internal Server Error. Other tools fail to exercise the \textsc{get} \smalltt{/playlists/{playlist\_id}/tracks} operation entirely and are hence unable to locate this error.

\begin{rqanswer}
AutoRestTest outperforms the other tools in terms of fault detection capability. It identifies a total of 42 instances of 500 Internal Server Errors, whereas ARAT-RL, EvoMaster, MoRest, and RESTler detected 33, 20, 20, and 14 errors, respectively.
\end{rqanswer}

\subsection{RQ3: Ablation Study}

\begin{table}[t]
\caption{Code coverage achieved by different tool variants.}
\centering
\label{tab:coverage_metrics}
\resizebox{\columnwidth}{!}{%
\begin{tabular}{lccc}
\toprule
 & Method & Line & Branch \\
\midrule
AutoRestTest & 58.3\% & 32.1\% & 58.3\% \\
1. Without LLM & 47.4\% (-10.9\%) & 19.3\% (-12.8\%) & 45.8\% (-12.5\%) \\
2. Without Learning & 45.6\% (-12.7\%) & 18.2\% (-13.9\%) & 45.8\% (-12.5\%) \\
3. Without SPDG & 46.7\% (-11.6\%) & 18.7\% (-13.4\%) & 47.6\% (-10.7\%) \\
\bottomrule
\end{tabular}
}
\end{table}

To understand the contribution of each component in AutoRestTest, we conducted an ablation study in which we removed specific elements of the approach: the LLM-based input generation, the agent learning step in MARL, and the SPDG. Because our tool heavily depends on agents, it was not feasible to create a reasonable tool without MARL entirely. Therefore, instead of removing all the agents, we only removed the temporal-difference Q-learning from the agents. Table~\ref{tab:coverage_metrics} presents the impact of these removals on method, line, and branch coverage.

Removing the temporal-difference Q-learning leads to the most significant decrease in overall metrics, dropping the coverage rates to 45.6\%, 18.2\%, and 45.9\% for method, line, and branch coverage. The learning step's contribution is crucial in optimizing the testing process through strategic exploration and exploitation of testing paths. Without the multi-agent learning step, the tool repeated the same requests and failed to properly update its agents with feedback.

The removal of the SPDG has the next most significant impact in terms of method and line coverage, dropping the performance significantly to 46.7\%, 18.7\%, and 47.6\% for method, line, and branch coverage. This result indicates that the SPDG plays a critical role in identifying dependencies and guiding test generation by reducing the search space, thus helping identify the dependent API operations.

Removing the LLM alone also leads to a substantial decrease in performance, with method, line, and branch coverage dropping to 47.4\%, 19.3\%, and 45.8\%. The primary reason for this difference is the LLM's ability to generate diverse and appropriate test inputs. These inputs are essential for exercising the API, as they help uncover various parameter and operation dependencies. Furthermore, operation and parameter dependencies cannot be accurately identified without resolving parameter constraints when they exist.

Notably, without the SPDG, AutoRestTest exercised only 5 operations for Spotify, whereas with the SPDG, it consistently covered 7 operations. 

\begin{rqanswer}
The ablation study shows that each component of AutoRestTest (LLM, MARL, and SPDG) contributes considerably to the effectiveness of the approach, and removing any component drops the coverage significantly. The decrease in method, line, and branch coverage ranges, in percentage points, between 10.9 and 12.7, 12.8 and 13.9, and 10.7 and 12.5, indicating that each component plays an important role.
\end{rqanswer}

\subsection{Threats to Validity}

Like for any empirical study, there are potential threats to the validity of our results. Regarding construct validity, the use of ChatGPT-3.5-Turbo as our LLM component~\cite{openai2023gpt4} introduces potential data leakage, as it may have been trained on API-related content, potentially affecting our results. While our ablation study demonstrates the LLM's importance, this limitation should be considered when interpreting our findings. Additionally, technical choices like the 20\% mutation rate~\cite{kim2023reinforcement} and ChatGPT's default parameters~\cite{kim2023leveraging} may affect result comparability.

REST APIs' inherently flaky behavior (e.g., due to network issues) introduces possible threats of internal validity. To mitigate this issue, we performed multiple test runs and averaged the results. We also carefully inspected our code and results to mitigate the risk of implementation errors.

The uneven quality of OpenAPI specifications~\cite{ruikai} can also affect the validity of our results. While this is an inherent and somehow unavoidable issue, we have left to future work the investigation of the impact of specification completeness and accuracy on AutoRestTest's performance. 

Finally, our selection of REST APIs benchmarks can affect external validity. While we used a diverse set of real-world APIs, AutoRestTest may perform differently on different benchmarks. The availability of our dataset and code will allow other researchers to validate and extend our evaluation.

\section{Related Work}
\label{sec:related}

\subsection{Automated REST API Testing}

Automated testing for REST APIs has employed various strategies. EvoMaster~\cite{arcuri2019restful} uses both white-box and black-box techniques and applies evolutionary algorithms to refine test cases, focusing on detecting server errors. RESTler~\cite{atlidakis2019restler} generates stateful tests by inferring producer-consumer dependencies, aiming to uncover server failures. Tools such as RestTestGen~\cite{Corradini2022} exploit data dependencies and utilize oracles to validate status codes and response schemas. MoRest~\cite{liu2022morest} adopts model-based testing to simulate user interactions and generate test cases, while RestCT~\cite{wu2022combinatorial} employs combinatorial testing techniques to explore parameter value combinations systematically. ARAT-RL~\cite{kim2023reinforcement} introduces reinforcement learning to adapt and refine API testing strategies based on real-time feedback. Techniques such as QuickREST~\cite{karlsson2020quickrest}, Schemathesis~\cite{zac2022schemathesis}, and RESTest~\cite{martin2021restest} use property-based testing and various oracles to ensure compliance with OpenAPI or GraphQL specifications. Tools such as Dredd~\cite{dredd}, fuzz-lightyear~\cite{fuzz-lightyear}, and Tcases~\cite{tcases} provide diverse testing capabilities, from validating responses against expected results to detecting vulnerabilities and validating Swagger-based specifications.

The closest approach to AutoRestTest that incorporates reinforcement learning is ARAT-RL. However, ARAT-RL does not employ a comprehensive model to represent operation dependencies, which reduces its effectiveness. Given the difficulties in embedding potential dependencies into the action space of an agent, ARAT-RL considers weighted probabilities using parameter frequency to guide potential dependencies. This can result in the inefficient prioritization of unrelated operations in subsequent requests. Unlike ARAT-RL, AutoRestTest uses the SPDG to narrow the dependency search.

MoRest, with its RESTful Property Graph (RPG), is the closest approach to AutoRestTest in terms of using a graph model. However, MoRest is incapable of progressively adapting its requests according to server feedback, rendering its dependency sequences less effective. 

Concurrently to this work, two additional REST API testing techniques were proposed: DeepREST~\cite{corradini2024rest} and LlamaRestTest~\cite{kim2025llamaresttesteffectiverestapi}. DeepREST uses deep reinforcement learning to discover implicit API constraints, employing a single agent that learns operation orderings through a reward mechanism. However, deep learning's black-box nature makes it difficult to track how and why specific testing decisions are made. In contrast, AutoRestTest's reward mechanism is more effective because uses multiple specialized agents and is able to track how each agent's decisions contribute to the overall testing strategy. Additionally, while DeepREST learns dependencies purely through trial and error, AutoRestTest reduces the search space upfront using the SPDG. LlamaRestTest fine-tunes small language models specifically for REST API testing tasks, using Llama models~\cite{dubey2024llama} to identify inter-parameter dependencies and generate valid inputs. LlamaRestTest mainly focuses on LLM-driven testing, whereas our approach uses LLMs only to generate parameter values while relying on the SPDG for efficient dependency identification and multi-agent reinforcement learning for dynamic optimization across all testing steps. We could not compare with either of these approaches in our evaluation because they had not yet been published when we performed this work.



\subsection{LLM-based REST API Testing and Analysis}

Recent advancements in LLMs have resulted in improved REST API testing and analysis approaches. NESTFUL~\cite{basu2024nestful} provides a benchmark for evaluating LLMs on nested API calls, revealing challenges in handling complex API interactions. RESTSpecIT~\cite{decrop2024you} demonstrates automated specification inference and black-box testing with minimal user input, while RestGPT~\cite{song2023restgpt} introduces coarse-to-fine online planning for improved task decomposition and API selection. While these approaches primarily focus on REST API analysis, AutoRestTest focuses on testing. RESTSpecIT includes some testing capabilities, but it is limited to basic parameter mutation and lacks advanced testing features such as operation/parameter dependency identification.

\subsection{Reinforcement Learning-Based Test Case Generation}
Recent studies have explored reinforcement learning for software testing, particularly for web and mobile applications. For instance, Zheng and colleagues proposed a curiosity-driven reinforcement learning approach for web client testing~\cite{zheng2021webrl}, and Pan and colleagues applied a similar technique for Android application testing~\cite{pan2020androidrl}. Other work includes QBE, a Q-learning-based exploration method for Android apps~\cite{koroglu2018qbe}, and AutoBlackTest, an automatic black-box testing approach for interactive applications~\cite{mariani2021autoblacktest}. Reinforcement learning has also been used specifically for Android GUI testing~\cite{adamo2018androidrl, vuong2018androidrl, koroglu2020androidrl}.
While these approaches use single-agent-based techniques, AutoRestTest decomposes the test-generation problem into multiple components (operation, parameter, value, dependency) and uses a multi-agent reinforcement algorithm to reward the different components' contributions in each reinforcement learning step. Additionally, AutoRestTest introduces the SPDG to reduce the search space of operation dependency.
 
\section{CONCLUSION AND FUTURE WORK}
\label{sec:conclusion}
In this paper, we introduced AutoRestTest, a new technique that leverages multi-agent reinforcement learning, the semantic property dependency graph, and large language models to enhance REST API testing.
By optimizing specialized agents for dependencies, operations, parameters, and values, AutoRestTest addresses the limitations of existing techniques and tools. Our evaluation on 12 state-of-the-art REST services shows that AutoRestTest can significantly outperform leading REST API testing tools in terms of code coverage, operation coverage, and fault detection. Furthermore, our ablation study confirms the individual contributions of the MARL, LLMs, and SPDG components to the tool's effectiveness. In future work, we will explore the dynamic adaptation of testing strategies, optimize performance and scalability (e.g., through fine-tuning LLMs), and develop more sophisticated fault-classification approaches.

\section*{Acknowledgments}
\begin{small}
  This work was partially supported by 
  NSF, under grant CCF-0725202 and
  DOE, under contract DE-FOA-0002460,
  and gifts from Facebook, Google, IBM Research, and Microsoft Research.
\end{small}

\balance
\bibliographystyle{IEEEtran}
\bibliography{paper}

\end{document}